\documentclass[12pt]{article}
\usepackage{amssymb,epsfig}

\newcommand{\ci}[1]{\cite{#1}}

\newcommand{\be}[1]{
\begin{equation}\label{#1}}
\newcommand{\ee}{\end{equation}}
\newcommand{\ba}[1]{
\begin{eqnarray}\label{#1}}
\newcommand{\ea}{\end{eqnarray}}
\newcommand{\baa}{\begin{eqnarray*}}
\newcommand{\btab}{\begin{tabular}}
\newcommand{\etab}{\end{tabular}}
\newcommand{\eaa}{\end{eqnarray*}}

\def \labeltest #1 {\label{#1}}
\newcommand\re[1]{(\ref{#1})}

\def\II{\hbox{{1}\kern-.25em\hbox{l}}}

\newcommand{\insertfig}[2]{\mbox{\epsfxsize=#1cm \epsfbox{#2.eps}}}

\begin{document}

\begin{titlepage}
\begin{flushright}
\begin{tabular}{l}
TPR--00--23\\
hep-ph/0012220
\end{tabular}
\end{flushright}
\vskip1cm
\begin{center}
  {\large \bf
  Hard Exclusive Production of Tensor Mesons
  \\}
\vspace{1cm}
{\sc V.M.~Braun {\rm and} N.~Kivel\footnote{on leave of absence from 
St.Petersburg
Nuclear Physics Institute, 188350, Gatchina, Russia }}

\vspace*{0.1cm} 
{\it
   Institut f\"ur Theoretische Physik, Universit\"at
   Regensburg, \\ D-93040 Regensburg, Germany
                 }\\[1.cm]
\vskip0.8cm
{\bf Abstract:\\[10pt]} \parbox[t]{\textwidth}{
 We point out that hard exclusive production of tensor mesons $f_2(1270)$ 
 with helicity $\lambda=\pm 2$ is dominated by the gluon  
 component in the meson wave function and can be used to determine 
 gluon admixture in tensor mesons in a theoretically clean manner.
 We present a detailed analysis of the tensor meson distribution amplitudes
 and calculate the transition form factor $\gamma+\gamma^*\to f_2(1270)$
 for one real and one virtual photon.}
\vskip1cm

\end{center}

\end{titlepage}

\newpage

{\large\bf 1.}~~Gluon admixture to meson wave functions is interesting for 
many reasons 
and has been subject to extensive and somewhat controversial discussion
over many years, see e.g. \ci{gluon}. 
The purpose of this letter is to point out that 
separation of quark and gluon contributions in hard processes that are
dominated by hadron wave functions at small transverse separations
has a different meaning compared to the quark model; in a suitably 
chosen reaction quark contribution can be down compared to the 
gluon contribution by a power of the momentum transfer.

In particular, we consider hard exclusive production of the tensor 
meson $f_2(1270)$ with the quantum numbers $J^{PC}=2^{++}$, $I^G=0^+$.
This state is non-exotic, and, in quark model, can be constructed 
either from a quark and an antiquark, or from a pair of gluons.  
To get the spin projection $s=2$ the quark and the antiquark have to be in
a P-wave state, while the gluons can be in $S$-wave.
On the light-cone, however, contribution of the orbital angular momentum 
is higher-twist and we will find that the form factor 
$\gamma+\gamma^*\to f_2(1270)$ corresponding to a pure helicity state 
$\lambda=\pm2$ is dominated at large photon virtualities 
by the gluon contribution.  

{\large\bf 2.}~~Distribution amplitudes of a tensor meson 
can be constructed in full analogy with those of vector and pseudoscalar 
mesons, see e.g. \cite{exclusive,ChZh,BBKT,BB99}. 
We consider matrix elements of bilocal 
quark-antiquark operators at a light-like separation $z_\mu$, $z^2=0$:
\ba{phi}
 \langle f_2(P,\lambda)|\bar \psi(z)\gamma_\mu \psi(-z)|0\rangle
 &=&
   f_q m^2 p_\mu \frac{e^{(\lambda)}_{\bullet\bullet}}
{p_{\scriptstyle\bullet}^2}
  \int\limits_{-1}^1 \! dt\,e^{itp_{\bullet}} \phi_q(t)
\nonumber\\
&+&f_q m^2 \left[e^{(\lambda)}_{\mu\bullet} - p_\mu 
\frac{e^{\scriptstyle\bullet\bullet}}{p_{\bullet}}\right]
\frac{1}{p_\bullet}
  \int\limits_{-1}^1 \! dt\,e^{itp_{\bullet}} g_v(t)
 +{\cal O}(z_\mu)\,,
\nonumber\\
 \langle f_2(P,\lambda)|\bar \psi(z)\gamma_\mu\gamma_5 \psi(-z)|0\rangle
 &=&
-i f_q m^2 \varepsilon^{\mu\nu\alpha\beta} \frac{z^\nu p^\alpha}{p_\bullet}
 \frac{e^{(\lambda)}_{\beta\bullet}}{p_\bullet}
  \int\limits_{-1}^1 \! dt\,e^{itp_{\scriptstyle\bullet}} g_a(t)\,,
\ea
where $\bar \psi \psi = \bar u u+\bar d d$. $P_\mu$ and $m=1270$~MeV
 are the $f_2$-meson momentum  and mass, respectively: $P^2=m^2$. 
Here and below we use a shorthand notation $p_\bullet = p_\mu z^\mu$, etc., 
and define the  light-like vector 
$p_\mu= P_\mu - z_\mu m^2/(2p_\bullet)$,
$p^2=0$. The polarization tensor $e^{(\lambda)}_{\alpha\beta}$ is 
symmetric and traceless, and satisfies the condition 
$e^{(\lambda)}_{\alpha\beta} P^\beta=0$.
Polarization sums can be calculated using 
\be{polarization}
\sum_\lambda e^{(\lambda)}_{\mu\nu}\left(e^{(\lambda)}_{\rho\sigma}\right)^\ast
  = \frac12 M_{\mu\rho} M_{\nu\sigma}+\frac12 M_{\mu\sigma} M_{\nu\rho}
  -\frac13 M_{\mu\nu} M_{\rho\sigma}\,,
\ee
where $M_{\mu\nu} = g_{\mu\nu} - P_\mu P_\nu/m^2$ and 
the normalization is such that 
$e^{(\lambda)}_{\mu\nu}\big(e^{(\lambda')}_{\mu\nu}\big)^\ast= 
\delta_{\lambda\lambda'}$.

The distribution amplitude $\phi_q(t) = -\phi_q(-t)$  is the leading twist-2 
distribution amplitude for the tensor mesons with helicity $\lambda=0$,
while $g_v(t)=-g_v(-t)$ and $g_a(t)=g_a(-t)$ 
correspond to the twist-three amplitudes 
with helicity $\lambda= 1$. The leading twist-2 
distribution amplitude for $\lambda= 1$ is given by a similar 
matrix element with the $\sigma_{\mu\bullet}$ matrix in between the 
quarks; it does not contribute to two-photon processes 
(if quark masses are neglected)  and will not be 
considered here.    

Normalization of the distribution amplitudes is chosen to be: 
\be{norm}
  \int\limits_{-1}^{1} dt\, t\,\phi_q(t) = 
  \int\limits_{-1}^{1} dt\, t\,g_v(t) = 1\,,\qquad 
  \int\limits_{-1}^{1} dt\, g_a(t)=0\,.
\ee
The constant $f_q$ is defined as the matrix element of the local operator
\be{norm1}
 \langle f_2(P,\lambda)|\bar \psi\gamma_\mu 
i\stackrel{\leftrightarrow}{D}_\nu\psi|0\rangle
 = f_q m^2 e^{(\lambda)}_{\mu\nu}
\ee
and was estimated using QCD sum rules \cite{AS82}:
\be{gf}
    f_q(\mu = 1~{\rm GeV}) \simeq 56~{\rm MeV}\,. 
\ee
The twist-three distribution amplitudes $g_v(t)$ and $g_a(t)$ receive 
a Wandzura-Wilczek type contribution that can be expressed in terms of the 
distribution function $\phi_q$ of the leading twist. A standard calculation
(cf. \cite{BBKT}) yields
\ba{WW}
    g_v^{WW}(t) &=& \int\limits_{-1}^t dw\, \frac{\phi_q(w)}{1-w}+
                  \int\limits_{t}^1 dw\, \frac{\phi_q(w)}{1+w}\,,
\nonumber\\
    g_a^{WW}(t) &=& \int\limits_{-1}^t dw\, \frac{\phi_q(w)}{1-w}-
                  \int\limits_{t}^1 dw\, \frac{\phi_q(w)}{1+w}\,.
\ea
Because of a large mass of the $f_2$-meson, we expect that the twist-3
distribution amplitudes are dominated by these ``kinematic'' contributions
whereas genuine twist-3 contributions of quark-gluon operators are relatively
small.

The conformal expansion of the leading-twist distribution amplitude 
$\phi_q(t)$ goes over the usual set of Gegenbauer polynomials 
$(1-t^2)\,C^{3/2}_n(t)$ with odd values of $n=1,3,\ldots$ because of the 
$C$-parity. The asymptotic wave function is, therefore
\be{phi-as}
    \phi_q^{\rm as}(t) = \frac{15}{4} t(1-t^2)
\ee 
and the corresponding expressions for the twist-3 distributions are 
\be{gav-as}
    g_v^{\rm as}(t) = \frac52 t^3\,,\qquad g_a^{\rm as}(t) = 
    \frac54 (3t^2-1)\,.
\ee 
In addition, there exist two different leading twist 
gluon distribution amplitudes
\ba{phi-g}
\langle f_2(P,\lambda)|
S_{\mu\nu}G^a_{\bullet\mu}(z)G^a_{\bullet\nu}(-z)|0\rangle
 &=&
 f^T_g e^{(\lambda)}_{\mu\nu} p_\bullet^2 \int\limits_{-1}^1 \! dt\,
e^{itp_{\bullet}} \phi_g^T(t) +{\cal O}(z_\mu, z_\nu)\,,
\nonumber\\
\langle f_2(P,\lambda)|
S_{\mu\nu}G^a_{\xi\mu}(z)G^a_{\xi\nu}(-z)|0\rangle
 &=&
 f^S_g e^{(\lambda)}_{\mu\nu} \int\limits_{-1}^1 \! dt\,
e^{itp_{\bullet}} \phi_g^S(t) +{\cal O}(z_\mu, z_\nu)\, .
\ea
Here $S_{\mu\nu}$ stands for the  symmetrisation in the two indices and removal of the 
trace: 
$ S_{\mu\nu}{\cal O}_{\mu\nu} = \frac12\, {\cal O}_{\mu\nu} +
\frac12\,{\cal O}_{\nu\mu}
 -\frac14\, g_{\mu\nu}{\cal O}_{\xi\xi}$.
The distribution amplitudes $\phi_g^T(t)$ and  $\phi_g^S(t)$ are 
both symmetric to the interchange 
of $t\leftrightarrow -t$ and describe the momentum fraction distribution
of the two gluons in the $f_2$-meson having the same and the opposite 
helicity, respectively. 
The asymptotic distributions at large scales are equal to 
\be{phi2as}
   \phi_g^{T,{\rm as}}(t) = \phi_2^{S,{\rm as}}(t)=  
   \frac{15}{16} (1-t^2)^2\,. 
\ee  
The constants $f^T_g$ and $f^S_g$ are defined through the matrix element of 
the local two-gluon operator:
\ba{GTnorm}
\langle f_2(P,\lambda)|
G^a_{\alpha\beta}(0)G^a_{\mu\nu}(0)|0\rangle &=&
f^T_g\biggl\{ 
\big[(P_\alpha P_\mu-\frac12 m^2g_{\alpha \mu})\, e^{(\lambda)}_{\beta\nu}-
 (\alpha\leftrightarrow\beta)\big]- (\mu\leftrightarrow\nu)\biggl\}
\nonumber\\
&+&\frac12 f^S_g m^2\biggl\{\big[g_{\alpha \mu}\,e^{(\lambda)}_{\beta\nu}
-  (\alpha\leftrightarrow\beta)\big]- (\mu\leftrightarrow\nu)\biggl\}.
\ea

The constant  $f^T_g$ is renormalized multiplicatively
\cite{LFBK85,Eric,JiH98}, while 
 $f^S_g$ is mixed with $f_q$ \cite{Grozin}:
\ba{GTscale}
f^T_g(Q^2)&=& f^T_g(\mu^2) L^{-1+6N_c/\beta_0}\,,
\nonumber\\
f_q(Q^2)&=& \frac{n_f}{n_f+4C_F}
\biggl(
f^S_g(\mu^2)+f_q(\mu^2)
\biggl)
-L^{\frac23(n_f+4C_F)/\beta_0}
\biggl(
f^S_g(\mu^2)-\frac{4C_F}{n_f}f_q(\mu^2)
\biggl),
\nonumber\\
f^S_g(Q^2)&=&\frac{4C_F}{n_f+4C_F}
\biggl(
f^S_g(\mu^2)+f_q(\mu^2)
\biggl)+
L^{\frac23(n_f+4C_F)/\beta_0}
\biggl(
f^S_g(\mu^2)-\frac{4C_F}{n_f}f_q(\mu^2)
\biggl),
\ea
where $L=\alpha_s(Q^2)/\alpha_s(\mu^2)$, $C_F=\frac{N_c^2-1}{2N_c}$ and  
$\beta_0 = 11/3 N_c -2/3 n_f$.
Note that the sum $f^S_g+f_q$ is scale-independent. 
We are not aware of any estimates of the numerical values of $f^T_g$, $f^S_g$.
The main goal of this letter is to point out that the constant $f^T_g$ 
can be measured in hard exclusive processes.

{\large\bf 3.}~~In the rest of the work we consider one particular hard 
reaction: the transition form factor 
$\gamma^*(q)+ \gamma(q')\rightarrow f_2(P)$
for one real $q'^2=0$ and one virtual $q^2=-Q^2$ photon.
The amplitude of this process is related to the matrix element
\be{Tdef}
T_{\mu\nu}=i\int d^4x \,e^{-ix(q-q')/2} 
\langle f_2(P,\lambda)|T\{ j_\mu(x/2)j_\nu(-x/2)\}|0\rangle,
\ee
where $j_\mu = e_u \bar u \gamma_\mu u + e_d \bar u \gamma_\mu u+\ldots$
is the electromagnetic current. 
The kinematics is illustrated in Fig.~1. 
\begin{figure}[t]
\begin{center}
\hspace{0cm}
\insertfig{3.5}{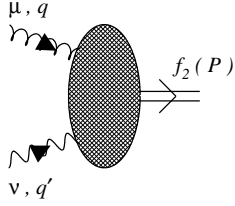}
\end{center}
\caption[dummy]{
\small The kinematics of the process 
$\gamma^*(q)+\gamma(q')\rightarrow f_2(P)$ 
\label{Fig1}
}
\end{figure}
Neglecting contributions that vanish after the multiplication by photon 
polarizations, the amplitude $T_{\mu\nu}$ can be parametrized in terms 
of the three invariant form factors
\ba{Tmunu}
T_{\mu\nu} &=& 
  -g_{\mu\nu}^\perp 
  e_{\alpha\beta}(q-q')^\alpha (q-q')^\beta \frac{m^2}{(2qq')^2} \,T_0(Q^2)+
\nonumber \\&&
  -g_{\nu\alpha}^\perp e_{\alpha\beta}(q-q')^\beta 
\biggl(q-q'\frac{q^2}{(qq')}\biggl)_\mu \frac{m^2}{(2qq')^2}\, T_1(Q^2)
\nonumber\\&&
+\Big(g_{\mu\alpha}^\perp g_{\nu\beta}^\perp+ 
g_{\mu\nu}^\perp\frac{(q-q')^2}{(2qq')^2}
(q-q')^\alpha (q-q')^\beta\Big) e^{\alpha\beta} T_2(Q^2)
\biggl],
\ea
 where we have introduced the metric tensor $g_{\mu\nu}^\perp$ 
 that is transverse to the photon momenta 
\be{gT}
g_{\mu\nu}^\perp=g_{\mu\nu} -
\frac1{(qq')}(q_\mu q'_\nu+q_\nu q'_\mu-\frac{q^2}{(qq')}q'_\mu q'_\nu)\,.
\ee
The form factors $T_0, T_1$ and $T_2$ correspond to the 
three possible helicity amplitudes 
\ba{Thelicity}
T_0: && \, \gamma^*(\pm 1)+\gamma(\pm 1)\rightarrow f_2(0)\, ,
\nonumber \\
T_1: &&\, \gamma^*(0)+\gamma(\pm 1)\rightarrow f_2(\mp 1)\, ,
\nonumber \\
T_2: &&\, \gamma^*(\pm 1)+\gamma(\mp 1)\rightarrow f_2(\pm 2)\, .
\ea 
In the limit $Q^2\rightarrow 0$ only $T_0$ and $T_2$ contribute and 
their values at $Q^2=0$ determine the two-photon decay width
\be{Ggg}
\Gamma_{\gamma\gamma}= \frac{(4\pi\alpha)^2}{60\pi m }
\Big( |T_0(0)|^2+2|T_2(0)|^2
\Big) ,
\ee
where $\alpha=1/137$ is the fine structure constant.

Experimentally it was found  that the contribution of the helicity zero
state to the two photon width $\Gamma_{\gamma\gamma}$ is very small
(see, for example, \ci{MARKII90, JADE90}). Using \cite{MARKII90}
\ba{Width}
\Gamma_{\gamma\gamma}=3.15\pm 0.04\pm 0.39~{\rm keV}, \qquad  
\frac{\Gamma_0(f_2\rightarrow \gamma\gamma)}
{\Gamma_2(f_2\rightarrow \gamma\gamma)}< 0.05\,\, (90\%\, \mbox{C.L.})
\ea
we obtain 
\be{Tnorm}
T_2(0)=(212 \pm 20)\, \mbox{MeV}, \qquad T_0(0)< 67\, \mbox{MeV}. 
\ee

Next, we consider the region where the virtuality of   
one of the photons is large:
\be{hard}
-q^2\equiv Q^2 \gg m^2,\, \, q'^2=0. 
\ee
In this kinematics, the form factors  
$\gamma^*(q)+ \gamma(q')\rightarrow f_2(P)$ can be calculated by the light-cone
expansion of the T-product of the electromagnetic currents in \re{Tdef}
\ba{OPE}
 i T\{ j_\mu(x)j_\nu(-x)\} &=& 
\frac{1}{(4\pi)^2 x^4}\sum e^2_q \Bigg\{\Big[
  \bar\psi(-x)\gamma_\nu\!\not\!x\gamma_\mu \psi(x)-
  \bar\psi(x)\gamma_\mu\!\not\!x\gamma_\nu \psi(-x)
                                             \Big] 
\\&& -i \frac{\alpha_s}{4\pi }
\int_{-1}^1\!\!\! du\!\!\int_{-1}^1\!\!\! dv\, (1+uv)
S_{\mu\nu} G_{x\mu}(ux)G_{x\nu}(-vx)\Bigg\}+ \dots\,,
\ea
where we have shown the contributions that will be relevant for what follows,
and $G_{x\mu} = x_\xi G_{\xi\mu}$, etc. The corresponding Feynman diagrams are 
shown in Fig.~2. Note that we take into account the contribution 
of the two-gluon operator with helicity $\lambda=2$ but do not consider
the gluon contribution with $\lambda=0$ that enters at the same level 
as the ${\cal O}(\alpha_s)$ corrections to the quark-antiquark operator.  
\begin{figure}[t]
\begin{center}
\hspace{0cm}
\insertfig{10}{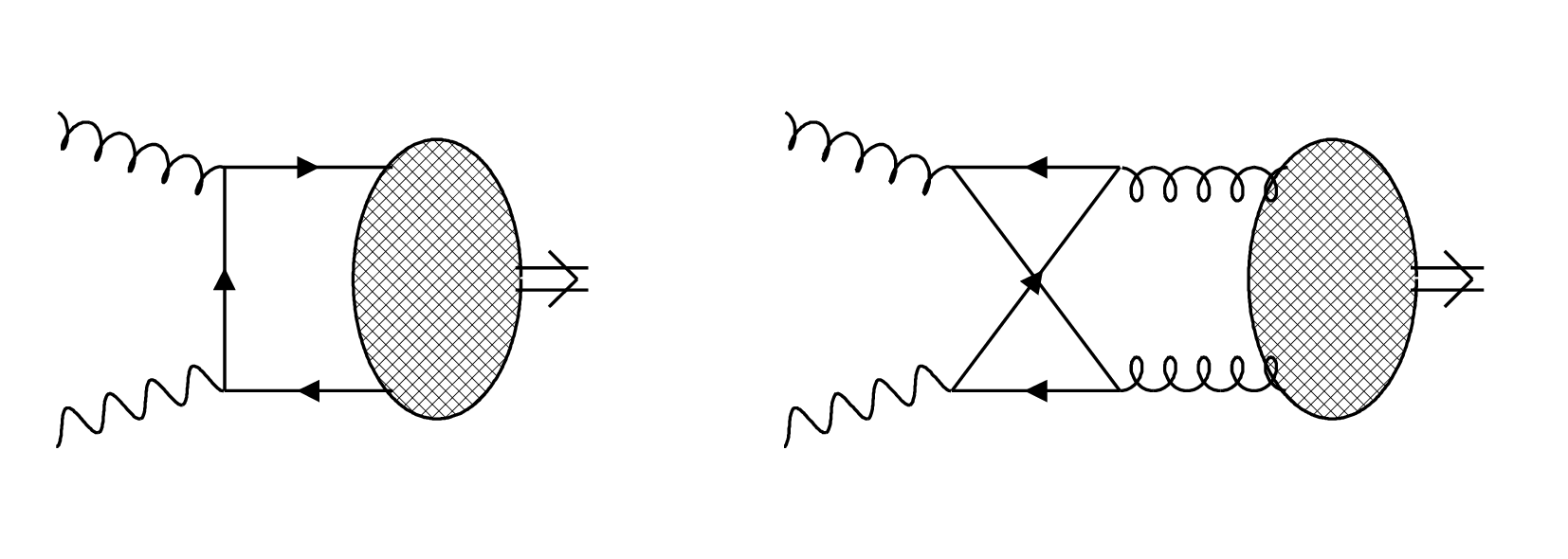}
\end{center}
\vspace{-0.5cm}
\caption[dummy]{
\small Leading-order contributions to 
 $\gamma^*(q)+ \gamma(q')\rightarrow f_2(P)$ in the large-$Q^2$ limit.  
\label{Fig2}
}
\end{figure}
In this paper we calculate the leading-twist contributions to the three 
form factors (helicity amplitudes) defined in \re{Tmunu} to the leading order 
in $\alpha_s$. For the most interesting case of the gluon-dominated amplitude 
$T_2$ we also keep the leading higher-twist correction.
A simple calculation gives:
\ba{T012}
T_0&=&\frac59 f_q  \int_{-1}^1\! \frac{tdt}{1-t^2}\,\phi_q (t)\,,
\nonumber\\
T_1&=&\frac59 f_q \, \int_{-1}^1\! dt \,\ln(1-t^2)\,g_a^{WW}(t)\,,
\nonumber\\
T_2&=& 2 \frac{\alpha_s}{\pi} \sum e^2_q\, f_g^T\,\int_{-1}^1
\,\frac{dt}{1-t^2}\, \phi_g^T(t)+ 
\nonumber \\
&+&\frac59 \frac{f_q m^2}{Q^2}\,\int_{-1}^1\! dt\, 
\ln\left(\frac{1+t}{1-t}\right)\,g_v^{WW}(t)\,,
\ea
where all distribution amplitudes and the coupling have to be taken 
at the scale $\mu^2=Q^2$.
Note that to ensure the gauge invariance the higher-twist distribution 
amplitudes must be taken, to our accuracy, 
in the Wandzura-Wilczek approximation. 
  
It is worthwhile to mention that hard exclusive $f_2$-meson 
production can be viewed as a particular case of the more general
hard exclusive two-pion production process considered in \ci{DPT}. 
Factorisation theorems derived for the two-pion production 
\cite{Freund} are valid for our case as well, and the 
results in \re{T012}, with the exception of the  
Wandzura-Wilzek contribution to $T_2$, can be  
extracted from the ones existing in the literature \ci{DPT,KMP,KM00} 
using a correspondence 
between the  two-pion soft matrix elements and the distribution amplitude
of a tensor  meson \ci{MPol}. 
In particular, the hard parton subprocesses and hence the coefficient 
functions are the same in both cases.

We expect that distribution amplitudes of the $f_2$-meson  are not far from 
their asymptotic form given in the text. On this assumption, which is 
certainly acceptable for an estimate, we obtain for large $Q^2$
\be{T2as}
T_2 \simeq \frac53\, \frac{\alpha_s}{\pi}\, f^T_g +\frac{50}{27}\, 
f_q\, \frac{m^2}{Q^2}\, , 
\ee  
where we have taken into account $u, d$ and $s$-quark contribution in the 
quark loop in the box diagram.
\begin{figure}[t]
\begin{center}
\hspace{0cm}
\insertfig{8}{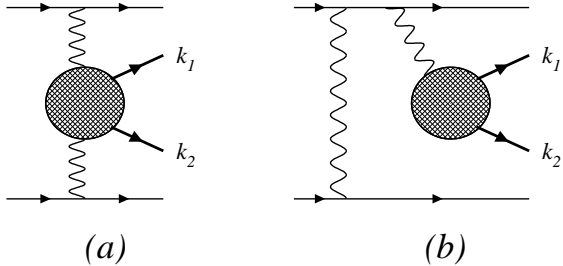}
\end{center}
\caption[dummy]{
\small The two subprocesses contributing to the reaction
$ee\rightarrow ee\pi^+\pi^-$ : $\gamma^*\gamma$ scattering (a) and
bremsstrahlung (b). 
\label{Fig3}
}
\end{figure}
For a realistic value $Q^2\sim 10$~GeV$^2$ the gluon contribution is comparable
to the subleading quark contribution if the  constants $f_g^T$ and $f_q $ 
are of the same order. It is interesting to note that $T_2(0)$ is much larger
than both $T_0(0)$ \re{Tnorm} and $f_q$ \re{gf}. This may indicate that 
$f_g^T$ is abnormally large.

Since $f_2$ decays in two pions with the branching ratio
about 95\%, one natural possibility to measure the form factors
is via the hard exclusive two-pion production 
$ee\to ee \pi\pi $, see Fig.~3a, that received a lot of attention 
recently \cite{DPT,Freund,KMP,KM00,Diehl}. This reaction is in fact observed
and constitites a background to the $\gamma^*\gamma\to \pi$ form factor 
measurements reported by CLEO \cite{Savinov}.
Let $k_1$  and $k_2$ be the momenta of the two pions 
in the final state. The tensor form factor of interest
$T_2(Q^2)$ is related to the two-pion helicity amplitude 
$A_{+-}$ (we use notations of Ref.~\cite{Diehl}, see Eq.~(75) there) 
in the  region $(k_1+k_2)^2\sim m^2$ as
\be{twopion}
A_{+-} = \varepsilon_+^\mu\varepsilon_-'^\nu (k_1-k_2)_\mu^\perp 
(k_1-k_2)_\nu^\perp
\frac{g_{f_2\pi\pi}}{m}\frac{T_2(Q^2)}{m^2-(k_1+k_2)^2},       
\ee
where $g_{f_2\pi\pi}$ is the corresponding decay constant:
\be{fpipi}
\langle \pi(k_1)\pi(k_2)|f_2(P,\lambda)\rangle=\frac{g_{f_2\pi\pi}}{m}
e_{\alpha\beta}^{(\lambda)}(k_1-k_2)^\alpha(k_1-k_2)^\beta
\ee
and $\varepsilon_+^\mu,\,\varepsilon_-'^\nu$ denote the transverse 
polarisation vectors of the virtual and the real photon, respectively. 
As was discussed  in Ref.~\cite{Diehl,KMP}  
the amplitude $A_{+-}$ can be separated  
using its nontrivial dependence on the azimuthal angle $\varphi$.  
Moreover, $A_{+-}$ is symmetric to the interchange of the 
pion momenta and can be measured by the interference with the 
(much larger) contribution of the bremsstrahlung
process  (see Fig.~3b) that is antisymmetric to the interchange 
of the pion momenta: In the difference of cross sections 
$\sigma(k_1,k_2)-\sigma(k_2,k_1)$
only the interference term survives. 
The relevant expressions 
for the physical cross sections  have been worked out in \cite{Diehl}.

{\large\bf 4.}~~To summarize, in this letter we point out a 
possibility to determine gluon admixture in tensor mesons using hard 
exclusive reactions. 
We construct the basic theoretical formalism and suggest a particular 
experimental setup where the relevant form factor can be measured.    
For simplicity, in this work we did not consider radiative corrections. 
NLO corrections to the coefficient functions have been calculated in
\cite{KMP} and NNLO evolution of the corresponding distributions
amplitudes can be found in  \cite{BelitskyS, BelitskyT}.

\subsection*{Acknowledgements}
This work was supported in part by the DFG, project Nr. 920585.

\end{document}